\documentclass[aps,prc,preprint,groupedaddress,showpacs,superscriptaddress,floatfix]{revtex4-1}

\usepackage{amsmath}
\usepackage{graphicx}
\usepackage{amssymb}
\usepackage{bm}
\usepackage{array}
\usepackage{hyperref}
\usepackage{slashed}
\usepackage{braket}

\begin{document}


\title{Generalized form factors of the nucleon in a light-cone spectator-diquark model}

\newcommand*{\PKU}{School of Physics and State Key Laboratory of Nuclear Physics and
Technology, Peking University, Beijing 100871,
China}\affiliation{\PKU}
\newcommand*{\CICQM}{Collaborative Innovation
Center of Quantum Matter, Beijing, China}\affiliation{\CICQM}
\newcommand*{\CHEP}{Center for High Energy
Physics, Peking University, Beijing 100871,
China}\affiliation{\CHEP}

\author{Tianbo Liu}\affiliation{\PKU}
\author{Bo-Qiang Ma}\email{mabq@pku.edu.cn}\affiliation{\PKU}\affiliation{\CICQM}\affiliation{\CHEP}

\date{\today}

\begin{abstract}
We investigate the generalized form factors of the nucleon in a light-cone spectator-diquark model. Compared to the form factors, the generalized form factors contain some more information of the structure of the nucleon. In our calculation, both the scalar and the axial-vector spectator-diquark are taken into account. As a relation between the spin in the instant form and that in the light-cone form, the Melosh-Wigner rotation effect is included for both the quark and the axial-vector diquark. We also provide numerical results from our model calculations, and the results are comparable with those from lattice QCD.
\end{abstract}

\pacs{12.39.Ki, 14.20.Dh, 12.38.Lg}

\maketitle

\section{Introduction}

Hadrons are bound states of the strong interaction which is described by the quantum chromodynamics (QCD) in the framework of the Yang-Mills gauge theory. One of the most challenging problems in particle physics is to understand the hadrons such as the proton and the neutron in terms of the quark and gluon degrees of freedom. Due to the nonperturbative nature of the QCD, it is complicated to calculate the properties and structures of hadrons, especially when the relativistic effects are taken into account. Although Euclidean lattice methods provide a very important first principle numerical simulation of the nonperturbative QCD~\cite{Wilson:1974sk,Kogut:1974ag}, it is limited by the enormous computational complexity, and dynamical observables in the Minkowski space-time are not directly obtained from the Euclidean lattice computations. The Dyson-Schwinger and Bethe-Salpeter methods are also powerful tools in studying the confinement and the dynamical chiral symmetry breaking~\cite{Cornwall:1981zr,Maris:2003vk}, but in practice the analyses are limited to the ladder approximation in the Landau gauge. Thus, at present, phenomenological models are necessary and helpful for us to provide a physical picture of the structure of hadrons.

The constituent quark model~\cite{GellMann:1964nj,Zweig:1981pd,Zweig:1964jf} and the parton model~\cite{Feynman:1969ej,Bjorken:1969ja} are proved successful in classifying the hadrons and explaining the experimental observations in high energy hadron scatterings. The light-cone form~\cite{Dirac:1949cp} of the QCD reconciles the covariant non-abelian quantum field theory and these models. In this form the fields are quantized at the fixed light-cone time $\tau=t+z$ instead of the usual time $t$, and its simple QCD vacuum structure allows an unambiguous definition of the constituents of a hadron. The hadronic properties are all encoded in the lignt-cone wave function in terms of their quark and gluon degrees of freedom, and the wave function is fully relativistic and frame-independent~\cite{Brodsky:1997de}.

The diquark was first mentioned in Gell-Mann's paper on quarks~\cite{GellMann:1964nj}. Then the quark-diquark model was introduced by Ida {\it et al.}~\cite{Ida:1966ev} and Lichtenberg {\it et al.}~\cite{Lichtenberg:1967zz,Lichtenberg:1968zz} to describe a baryon as a two-body system of a quark and a diquark. In this model, many baryonic properties were calculated, such as the baryon spectra~\cite{Lichtenberg:1981pp,Ono:1972ge,Ono:1973fj,Goldstein:1979wba}, magnetic moments~\cite{Isgur:1977mx,Isgur:1979ee}, neutron charge radius~\cite{Carlitz:1977bd,Isgur:1980hh,Dziembowski:1979tr}, baryon decays~\cite{Ono:1972ge,Ono:1973fj,Hayashi:1978he,Shito:1980mx,Bediaga:1984ib,Dosch:1988hu,Efimov:1990uz} and structure functions~\cite{Schmidt:1977hs,FernandezPacheco:1977je,Donnachie:1980iy,Pavkovic:1976re,Pavkovic:1976rf,Abbott:1979je,Abbott:1979yb,Close:1980rk}. However, the original quark-diquark model is not a fully relativistic model, and the physical picture of the diquark as a two-quark tightly bound state is unnatural. In the high energy scattering experiments, especially the lepton nucleon scatterings, the nucleon can be regarded as an active quark which is struck by the virtual photon and a spectator-diquark which does not couple to the virtual photon directly~\cite{Feynman:1973xc}. From this point of view, the diquark is the remaining constituents of the nucleon to provide the other quantum numbers, and some nonperturbative effects are taken into account by the mass of the spectator effectively. In order to provide a relativistic description, the Wigner rotation~\cite{Wigner:1939cj} or the Melosh transformation~\cite{Melosh:1974cu,Buccella:1974bz}, which relate the spinors in the instant form to those in the light-cone form, should be taken into account. This effect plays an important role in understanding the ``proton spin puzzle''~\cite{Ma:1991xq,Ma:1992sj}. This light-cone spectator-diquark model were applied to calculate the electromagnetic and axial form factors, transition form factors, structure functions and transverse momentum dependent parton distributions (TMDs)~\cite{Ma:2002ir,Ma:2002xu,Ma:1996np,Ma:1997gy,Ma:1997pm,Ma:1998ar,Ma:2000cg,Ma:2000ip,Ma:2001rm,Lu:2004au,She:2008tu,Bacchetta:2008af,She:2009jq,Bacchetta:2010si,Zhu:2011zza,Zhu:2011ir,Lu:2011pt,Zhu:2011ym,Lu:2012ez,Lu:2012gu}, and the results matched the data well. Therefore, it is necessary to extand its application to other physical observables as a further test and to help us understand the structure of nucleons more clearly.

In this paper, we calculate the generalized form factors of the nucleon in the light-cone spectator-diquark model. Compared to the ordinary form factors, they contain some more information about the internal structure of the nucleon. We also provide numerical results of our calculations, and the results are comparable with the lattice QCD data. The paper is organized as follows. In Sec. II, we briefly review the light-cone spectator-diquark model. Then we calculate the generalized form factors of the nucleon in Sec. III, and provide our numerical predictions in Sec. IV. A summary is contained in the last section.

\section{The light-cone spectator diquark model}

A free hadron state is an eigenstate of the Lorentz invariant light-cone hamiltonian $H_{LC}=P^+P^--\bm{P}_\perp^2$ with the invariant mass square as the eigenvalue. The hadron state can be expanded on the complete basis of free multiparticle Fock states which are quantized at the fixed light-cone time $\tau=t+z$~\cite{Brodsky:1997de},
\begin{eqnarray}
\left|\psi_H:P^+,\bm{P}_\perp,S_z\right\rangle=\sum_{n,\{\lambda_i\}}\prod_{i=1}^N\int\frac{dx_id^2\bm{k}_{\perp i}}{2\sqrt{x_i}(2\pi)^3}(16\pi^3)\delta\left(1-\sum_{j=1}^N x_j\right)\delta^{(2)}\left(\sum_{j=1}^N \bm{k}_{\perp j}\right)\nonumber\\
\psi_{n/H}(\{x_j\},\{\bm{k}_{\perp j}\},\{\lambda_j\})\left|n:\{x_iP^+\},\{x_i\bm{P}_\perp+\bm{k}_{\perp i}\},\{\lambda_i\}\right\rangle,
\end{eqnarray}
where $N$ is the number of the components in the state $|n\rangle$, $x_i=k^+_i/P^+$ is the light-cone momentum fraction of the $i$th component and $\bm{k}_{\perp i}$, $\lambda_i$ are its intrinsic transverse momentum and helicity. The projection of hadron eigenstate $|\psi_H\rangle$ on the Fock state $|n\rangle$ is the light-cone wave function, $\psi_{n/H}$, which is frame-independent.

The hadron eigenstate is normalized as
\begin{equation}
\left\langle\psi_H:P^+,\bm{P}_\perp,S_z|\psi_H:P'^+,\bm{P}'_\perp,S'_z\right\rangle=2P^+(2\pi)^3\delta(P^+-P'^+)\delta^{(2)}(\bm{P}_\perp-\bm{P}'_\perp)\delta_{S_zS'_z},
\end{equation}
and correspondingly the light-cone wave function is normalized as
\begin{equation}
\sum_{n,\{\lambda_i\}}\prod_{i=1}^N\int\frac{dx_id^2\bm{k}_{\perp i}}{2(2\pi)^3}(16\pi^3)\delta\left(1-\sum_{j=1}^N x_j\right)\delta^{(2)}\left(\sum_{j=1}^N \bm{k}_{\perp j}\right)|\psi_{n/H}(\{x_j\},\{\bm{k}_{\perp j}\},\{\lambda_j\})|^2=1.
\end{equation}
The one-particle state is defined by $|p\rangle=\sqrt{2p^+}a^\dagger(p)|0\rangle$ with (anti)commutation relations
\begin{equation}
[a(p),a^\dagger(p')]=\{b(p),b^\dagger(p')\}=(2\pi)^3\delta(p^+-p'^+)\delta^{(2)}(\bm{p}_\perp-\bm{p}'_\perp),
\end{equation}
where the $a(p)$, $a^\dagger(p)$, $b(p)$ and $b^\dagger(p)$ are annihilation and creation operators for bosons and fermions respectively.

For a nucleon the leading term in the Fock states expansion is the valence quarks state $|qqq\rangle$. In the impulse approximation, a single constituent quark is struck by the lepton and the remain part is regarded as a spectator-diquark which does not interact with the lepton. In fact, some nonperturbative effects between the spectator quarks and gluons from higher Fock states, $|qqqg\rangle$, $|qqqq\bar{q}\rangle$, $\cdots$, can be effectively absorbed into the mass of the spectator-diquark. Thus a spin one-half nucleon in the light-cone spectator-diquark model is effectively expressed as
\begin{equation}
\left|\psi_N\right\rangle=\sin\theta\,\phi_S|qS\rangle+\cos\theta\,\phi_V|qV\rangle,
\end{equation}
where $S$ and $V$ represent the scalar and axial-vector diquark respectively, $\phi_S$ and $\phi_V$ are the momentum space light-cone wave functions and $\theta$ is a angle to describe the $SU(6)$ spin-flavor symmetry breaking. In this paper, we choose the $SU(6)$ symmetry case $\theta=\pi/4$.

To write down the spin space wave function, we start from the $SU(6)$ quark model in the instant form. The quark-spectator-diquark states for the proton are written as
\begin{equation}
\begin{split}
|qS\rangle^{\uparrow/\downarrow}=&u_T^{\uparrow/\downarrow}S(ud),\\
|qV\rangle^{\uparrow/\downarrow}=&\pm\frac{1}{3}[u_T^{\uparrow/\downarrow}V_T^0(ud)-\sqrt{2}u_T^{\downarrow/\uparrow}V_T^{\pm1}(ud)\\
&-\sqrt{2}d_T^{\uparrow/\downarrow}V_T^0(uu)+2d_T^{\downarrow/\uparrow}V_T^{\pm1}(uu)].
\end{split}
\end{equation}
Then we transform the instant form spinors to the light-lone form through the Melosh-Wigner rotation~\cite{Wigner:1939cj,Melosh:1974cu},
\begin{equation}
\begin{split}
\chi_T^\uparrow&=w[(k^++m)\chi_F^\uparrow-(k^1+ik^2)\chi_F^\downarrow],\\
\chi_T^\downarrow&=w[(k^++m)\chi_F^\downarrow+(k^1-ik^2)\chi_F^\uparrow],
\end{split}
\end{equation}
where $w=1/\sqrt{2k^+(k^0+m)}$ and the subscripts $T$ and $F$ represent the instant form and light-cone form spinors respectively. This result is in agreement with that of Lepage and Brodsky~\cite{Lepage:1980fj}. The scalar spectator-diquark does not transform since it has spin-zero. For the axial-vector spectator diquark, the Melosh transformation for a spin-1 particle is written as~\cite{Ahluwalia:1993xa}
\begin{equation}
\begin{split}
V_T^{+1}=&w^2[(k^++m)^2V_F^{+1}-\sqrt{2}(k^++m)(k^1+ik^2)V_F^0+(k^1+ik^2)^2V_F^{-1}],\\
V_T^{0\ }=&w^2[\sqrt{2}(k^++m)(k^1-ik^2)V_F^{+1}\\
&+2(k^+(k^0+m)-(k^1-ik^2)(k^1+ik^2))V_F^0-\sqrt{2}(k^++m)(k^1+ik^2)V_F^{-1}],\\
V_T^{-1}=&w^2[(k^1-ik^2)^2V_F^{+1}+\sqrt{2}(k^++m)(k^1-ik^2)V_F^0+(k^++m)^2V_F^{-1}].
\end{split}
\end{equation}
For the momentum space light-cone wave function, we assume the Brodsky-Huang-Lepage (BHL) prescription~\cite{Brodsky:1980vj,Brodsky:1981jv,Brodsky:1982nx},
\begin{equation}
\phi_D(x,\bm{k}_\perp)=A_D\exp\left\{-\frac{1}{8\beta_D^2}\left[\frac{m_q^2+\bm{k}_\perp^2}{x}+\frac{m_D^2+\bm{k}_\perp^2}{1-x}\right]\right\},
\end{equation}
where the subscript $D$ represents the diquark with $S$ for the scalar and $V$ for the axial-vector, $m_q$ and $m_D$ are the masses of the quark and the spectator-diquark, $\beta_D$ is the harmonic oscillator scale parameter and $A_D$ is the normalization factor.

\section{The generalized form factors of the nucleon}

The QCD Lagrangian density is
\begin{equation}
\mathcal{L}_{QCD}=\bar{\psi}(i\gamma^\mu \mathcal{D}_\mu-m)\psi-\frac{1}{4}F^{a\mu\nu}F^a_{\mu\nu},
\end{equation}
where $\mathcal{D}_\mu=\partial_\mu-igA^a_\mu t^a$ is the covariant derivative and $F^a_{\mu\nu}=\partial_\mu A_\nu-\partial_\nu A_\mu+gf^{abc}A^b_\mu A^c_\nu$ is the field strength tensor with $[t^a,t^b]=if^{abc}t^c$ and $a$, $b$, $c$ are $SU(3)$ color octet indices. There are six towers of twist-two operators which form totally symmetric representations of the Lorentz group~\cite{Boffi:2007yc}:
\begin{eqnarray}
\mathcal{O}_q^{\mu\mu_1\cdots\mu_{n-1}}&=&\bar{\psi}\gamma^{(\mu}i\overleftrightarrow{\mathcal{D}}^{\mu_1}\cdots i\overleftrightarrow{\mathcal{D}}^{\mu_{n-1})}\psi\label{gffv},\\
\tilde{\mathcal{O}}_q^{\mu\mu_1\cdots\mu_{n-1}}&=&\bar{\psi}\gamma^{(\mu}\gamma_5i\overleftrightarrow{\mathcal{D}}^{\mu_1}\cdots i\overleftrightarrow{\mathcal{D}}^{\mu_{n-1})}\psi\label{gffa},\\
\mathcal{O}_{qT}^{\mu\nu\mu_1\cdots\mu_{n-1}}&=&\bar{\psi}i\sigma^{\mu(\nu}i\overleftrightarrow{\mathcal{D}}^{\mu_1}\cdots i\overleftrightarrow{\mathcal{D}}^{\mu_{n-1})}\psi\label{gfft},\\
\mathcal{O}_g^{\mu\mu_1\cdots\mu_{n-1}\nu}&=&F^{(\mu\alpha}i\overleftrightarrow{\mathcal{D}}^{\mu_1}\cdots i\overleftrightarrow{\mathcal{D}}^{\mu_{n-1}}F_\alpha^{\ \nu)},\\
\tilde{\mathcal{O}}_g^{\mu\mu_1\cdots\mu_{n-1}\nu}&=&-iF^{(\mu\alpha}i\overleftrightarrow{\mathcal{D}}^{\mu_1}\cdots i\overleftrightarrow{\mathcal{D}}^{\mu_{n-1}}\tilde{F}_\alpha^{\ \nu)},\\
\mathcal{O}_{gT}^{\mu\mu_1\cdots\mu_{n-1}\nu\alpha\beta}&=&F^{(\mu\alpha}i\overleftrightarrow{\mathcal{D}}^{\mu_1}\cdots i\overleftrightarrow{\mathcal{D}}^{\mu_{n-1}}F^{\nu)\beta},
\end{eqnarray}
where $\tilde{F}^{\mu\nu}$ is the dual field strength tensor and $\overleftrightarrow{\mathcal{D}}_\mu=(\overrightarrow{\mathcal{D}}_\mu-\overleftarrow{\mathcal{D}}_\mu)/2$.

In this work, we focus on the quark part operators, {\it i.e.} Eqs.~(\ref{gffv}-\ref{gfft}). For $n=1$, these operators reduce to the vector, axial-vector and tensor currents, and the corresponding electromagnetic, axial-vector and tensor form factors can be defined. For $n=2$, the generalized form factors of a spin one-half nucleon are defined as
\begin{eqnarray}
\langle P'S'|\bar{\psi}\gamma^{(\mu}i\overleftrightarrow{\mathcal{D}}^{\nu)}\psi(0)|PS\rangle\label{gff2v}
&=&\bar{u}(P',S')\bigg[\gamma^{(\mu}\overline{P}^{\nu)}A(Q^2)-\frac{q_\alpha}{2M}i\sigma^{\alpha(\mu}\overline{P}^{\nu)}B(Q^2)\\
&&\quad +\frac{1}{M}(q^\mu q^\nu-g^{\mu\nu}q^2)C(Q^2)+g^{\mu\nu}M\bar{c}(Q^2)\bigg]u(p,S),\nonumber\\
\langle P'S'|\bar{\psi}\gamma^{(\mu}\gamma_5i\overleftrightarrow{\mathcal{D}}^{\nu)}\psi(0)|PS\rangle\label{gff2a}
&=&\bar{u}(P',S')\bigg[\gamma^{(\mu}\gamma_5\overline{P}^{\nu)}\tilde{A}(Q^2)+\frac{1}{2M}q^\mu q^\nu\gamma_5\tilde{B}(Q^2)\bigg]u(P,S),\\
\langle P'S'|\bar{\psi}i\sigma^{\mu(\nu}i\overleftrightarrow{\mathcal{D}}^{\rho)}\psi(0)|PS\rangle\label{gff2t}
&=&\mathop{\bm{A}}_{\mu,\nu}\bar{u}(P',S')\bigg[i\sigma^{\mu(\nu}\overline{P}^{\rho)}A_T(Q^2)+\frac{1}{M^2}\overline{P}^\mu q^{(\nu}\overline{P}^{\rho)}\tilde{A}_T(Q^2)\\
&&\quad +\frac{1}{2M}\gamma^\mu q^{(\nu}\overline{P}^{\rho)}B_T(Q^2)+\frac{1}{M}\gamma^\mu\overline{P}^{(\nu}q^{\rho)}\tilde{B}_T(Q^2)\bigg]u(P,S),\nonumber
\end{eqnarray}
where $\overline{P}=(P+P')/2$ is the average nucleon four-momentum, $q=P'-P$ is the transferred four-momentum with $Q^2=-q^2$, $u(P,S)$ is the Dirac spinor, and $\mathop{\bm{A}}$ represents the antisymmetrization of the indices $\mu$ and $\nu$.

In the light-cone gauge $A^+=0$, no ghosts exist and the gluon has physical spin projections $J^z=\pm 1$. The plus component of the covariant derivative $\mathcal{D}^+$ in this gauge is exactly the ordinary derivative $\partial^+$. Then we expand the Dirac field operator as
\begin{equation}
\psi(x)=\sum_\lambda\int\frac{d\ell^+}{\sqrt{2\ell^+}}\frac{d^2\bm{\ell}_\perp}{(2\pi)^3}[b_\lambda(\ell)u(\ell,\lambda)e^{-i\ell\cdot x}+d_\lambda^\dagger(\ell)v(\ell,\lambda)e^{i\ell\cdot x}],
\end{equation}
and take the plus-plus component of the operators in Eqs.~(\ref{gff2v}) and (\ref{gff2a}):
\begin{eqnarray}\label{operator1}
\mathcal{O}^{++}(0)&=&\left.i\bar{\psi}(x)\gamma^+\overleftrightarrow{\partial}^+\psi(x)\right|_{x=0}\\
&=&\frac{1}{2}\int\frac{d\ell'^+d^2\bm{\ell}'_\perp}{(2\pi)^3}\frac{d\ell^+d^2\bm{\ell}_\perp}{(2\pi)^3}(\ell'^++\ell^+)\sum_\lambda[b^\dagger_\lambda(\ell')b_\lambda(\ell)+d^\dagger_\lambda(\ell)d_\lambda(\ell')],\nonumber\\
\tilde{\mathcal{O}}^{++}(0)&=&\left.i\bar{\psi}(x)\gamma^+\gamma_5\overleftrightarrow{\partial}^+\psi(x)\right|_{x=0}\\
&=&\frac{1}{2}\int\frac{d\ell'^+d^2\bm{\ell}'_\perp}{(2\pi)^3}\frac{d\ell^+d^2\bm{\ell}_\perp}{(2\pi)^3}(\ell'^++\ell^+)\nonumber\\&&\times[b^\dagger_\uparrow(\ell')b_\uparrow(\ell)-b^\dagger_\downarrow(\ell')b_\downarrow(\ell)-d^\dagger_\uparrow(\ell)d_\uparrow(\ell')+d^\dagger_\downarrow(\ell)d_\downarrow(\ell')].\nonumber
\end{eqnarray}
Here we use the Lepage-Brodsky conventions for the properties of the light-cone spinors~\cite{Lepage:1980fj}. For the operator in Eq.~(\ref{gff2t}), the plus-plus-plus component vanishes due to the antisymmetry of the first two indices. However the covariant derivative only appears with the latter two indices. Therefore we can choose a transverse indice for the first one. Then the operator is expressed as
\begin{eqnarray}\label{operator2}
\mathcal{O}_T^{1++}(0)&=&\left.\bar{\psi}(x)i\sigma^{1+}i\overleftrightarrow{\partial}^+\psi(x)\right|_{x=0}\\
&=&-\frac{1}{2}\int\frac{d\ell'^+d^2\bm{\ell}'_\perp}{(2\pi)^3}\frac{d\ell^+d^2\bm{\ell}_\perp}{(2\pi)^3}(\ell'^++\ell^+)\nonumber\\
&&\times[b^\dagger_\uparrow(\ell')b_\downarrow(\ell)-b^\dagger_\downarrow(\ell')b_\uparrow(\ell)+d^\dagger_\uparrow(\ell)d_\downarrow(\ell')-d^\dagger_\downarrow(\ell)d_\uparrow(\ell')]\nonumber
\end{eqnarray}

To calculate the generalized form factors, we choose the frame as
\begin{eqnarray}
P&=&(P^+,\frac{M^2}{P^+},0,0),\\
q&=&(0,\frac{2q\cdot P}{P^+},q^1,q^2),
\end{eqnarray}
where $2q\cdot P=-q^2=Q^2$. Then we can calculate the generalized form factors through
\begin{eqnarray}\label{expr1}
\langle P',\uparrow|\mathcal{O}^{++}(0)|P,\uparrow\rangle&=&2(P^+)^2A(Q^2),\\
\langle P',\uparrow|\mathcal{O}^{++}(0)|P,\downarrow\rangle&=&2(P^+)^2\frac{-(q^1-iq^2)}{2M}B(Q^2),\\
\langle P',\uparrow|\tilde{\mathcal{O}}^{++}(0)|P,\uparrow\rangle&=&2(P^+)^2\tilde{A}(Q^2),\\
\langle P',\uparrow|\mathcal{O}_T^{1++}(0)|P,\uparrow\rangle&=&-2(P^+)^2\frac{q^1}{4M}[\tilde{A}_T(Q^2)+\frac{1}{2}B_T(Q^2)+\tilde{B}_T(Q^2)],\\
\langle P',\uparrow|\mathcal{O}_T^{1++}(0)|P,\downarrow\rangle&=&-2(P^+)^2A_T(Q^2)-2(P^+)^2\frac{q^1(q^1-iq^2)}{8M^2}\tilde{A}_T(Q^2).\label{expr2}
\end{eqnarray}

As introduced in Sec.~II, the nucleon spin state can be written as a superposition of a series of quark-spectator-diquark states:
\begin{equation}\label{state}
|P,\uparrow\rangle=\sum_{q,D,\lambda,\Lambda}\int\frac{dx}{2\sqrt{x}}\frac{d^2\bm{k}_\perp}{(2\pi)^3}\frac{dx_D}{2\sqrt{x_D}}\frac{d^2\bm{k}_{D\perp}}{(2\pi)^3}16\pi^3\delta(1-x-x_D)\delta^{(2)}(\bm{P}_\perp-\bm{k}_\perp-\bm{k}_{D\perp})C^{qD}_{\lambda\Lambda}(k)|q_\lambda(k)D_\Lambda(k_D)\rangle,
\end{equation}
where $q$ is the flavor of the quark, $D$ is the kind of the diquark, and the state $|q_\lambda(k)D_\Lambda(k_D)\rangle$ can be expressed with the creation operators as
\begin{equation}
|q_\lambda(k)D_\Lambda(k_D)\rangle=\sqrt{2k^+}\sqrt{2k_D^+}\,b_\lambda^\dagger(k)a_\Lambda^\dagger(k_D)|0\rangle.
\end{equation}
The subscripts $\lambda$ and $\Lambda$ represent the spins of the quark and the spectator-diquark respectively, and they are defined in the light-cone form. The momentum of the quark is
\begin{equation}
k=(k^+,\frac{m^2+\bm{k}_\perp^2}{k^+},\bm{k}_\perp)=(xP^+,\frac{m^2+\bm{k}_\perp^2}{xP^+},k^1,k^2),
\end{equation}
and the momentum of the spectator-diquark is
\begin{equation}
k_D=(k_D^+,\frac{m_D^2+\bm{k}_{D\perp}^2}{k_D^+},\bm{k}_{D\perp})=((1-x)P^+,\frac{m_D^2+\bm{k}_\perp^2}{(1-x)P^+},-k^1,-k^2).
\end{equation}
Then the coefficients $C^{qD}_{\lambda\Lambda}$ derived with the Melosh-Wigner rotations are
\begin{eqnarray}
C^{uS}_{\uparrow0}(k)&=&\frac{1}{\sqrt{2}}\frac{m+xP^+}{\sqrt{(m+xP^+)^2+\bm{k}_\perp^2}}\phi_S(x,\bm{k}_\perp),\\
C^{uS}_{\downarrow0}(k)&=&-\frac{1}{\sqrt{2}}\frac{k^1+ik^2}{\sqrt{(m+xP^+)^2+\bm{k}_\perp^2}}\phi_S(x,\bm{k}_\perp),\\
C^{uV}_{\uparrow+1}(k)&=&-\frac{\sqrt{2}(k^1-ik^2)(m+m_V+P^+)(m_V-xP^++P^+)}{3\sqrt{2}\sqrt{(m+xP^+)^2+\bm{k}_\perp^2}(\bm{k}_\perp^2+(m_V-xP^++P^+)^2)}\phi_V(x,\bm{k}_\perp),\\
C^{uV}_{\uparrow0}(k)&=&\frac{(m+xP^+)(m_V-xP^++P^+)^2-\bm{k}_\perp^2(m+2m_V+2P^+-xP^+)}{3\sqrt{2}\sqrt{(m+xP^+)^2+\bm{k}_\perp^2}(\bm{k}_\perp^2+(m_V-xP^++P^+)^2)}\phi_V(x,\bm{k}_\perp),\\
C^{uV}_{\uparrow-1}(k)&=&-\frac{\sqrt{2}(k^1+ik^2)(\bm{k}_\perp^2-(m+xP^+)(m_V-xP^++P^+))}{3\sqrt{2}\sqrt{(m+xP^+)^2+\bm{k}_\perp^2}(\bm{k}_\perp^2+(m_V-xP^++P^+)^2)}\phi_V(x,\bm{k}_\perp),\\
C^{uV}_{\downarrow+1}(k)&=&-\frac{\sqrt{2}(m_V-xP^++P^+)((m+xP^+)(m_V-xP^++P^+)-\bm{k}_\perp^2)}{3\sqrt{2}\sqrt{(m+xP^+)^2+\bm{k}_\perp^2}(\bm{k}_\perp^2+(m_V-xP^++P^+)^2)}\phi_V(x,\bm{k}_\perp),\\
C^{uV}_{\downarrow0}(k)&=&\frac{(k^1+ik^2)(\bm{k}_\perp^2-(m_V-xP^++P^+)(2m+m_V+xP^++P^+))}{3\sqrt{2}\sqrt{(m+xP^+)^2+\bm{k}_\perp^2}(\bm{k}_\perp^2+(m_V-xP^++P^+)^2)}\phi_V(x,\bm{k}_\perp),\\
C^{uV}_{\downarrow-1}(k)&=&-\frac{\sqrt{2}(k^1+ik^2)^2(m+m_V+P^+)}{3\sqrt{2}\sqrt{(m+xP^+)^2+\bm{k}_\perp^2}(\bm{k}_\perp^2+(m_V-xP^++P^+)^2)}\phi_V(x,\bm{k}_\perp),\\
C^{dS}_{\lambda0}(k)&=&0,\\
C^{dV}_{\lambda\Lambda}(k)&=&-\sqrt{2}C^{uV}_{\lambda\Lambda}(k).
\end{eqnarray}
The expansions of the other proton spin state $|P,\downarrow\rangle$ and the final state $\langle P',\uparrow|$ are similar. Substituting the operators (\ref{operator1})-(\ref{operator2}) and the proton state expansions (\ref{state}) into Eqs.~(\ref{expr1})-(\ref{expr2}), we can obtain the expressions of the generalized form factors in the light-cone spectator-diquark model.

\section{Numerical results and discussions}

\begin{figure}
\includegraphics[width=0.5\textwidth]{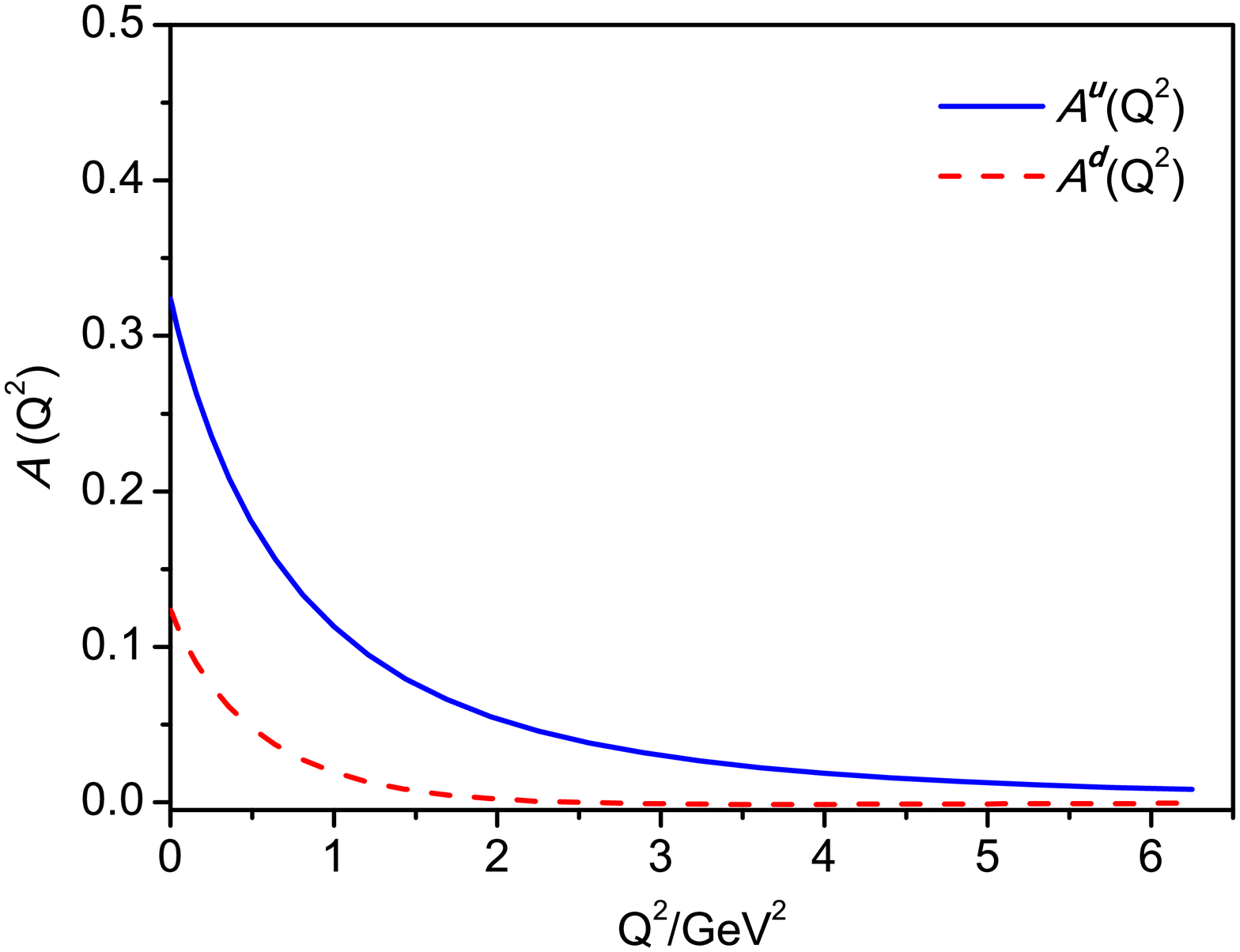}
\includegraphics[width=0.5\textwidth]{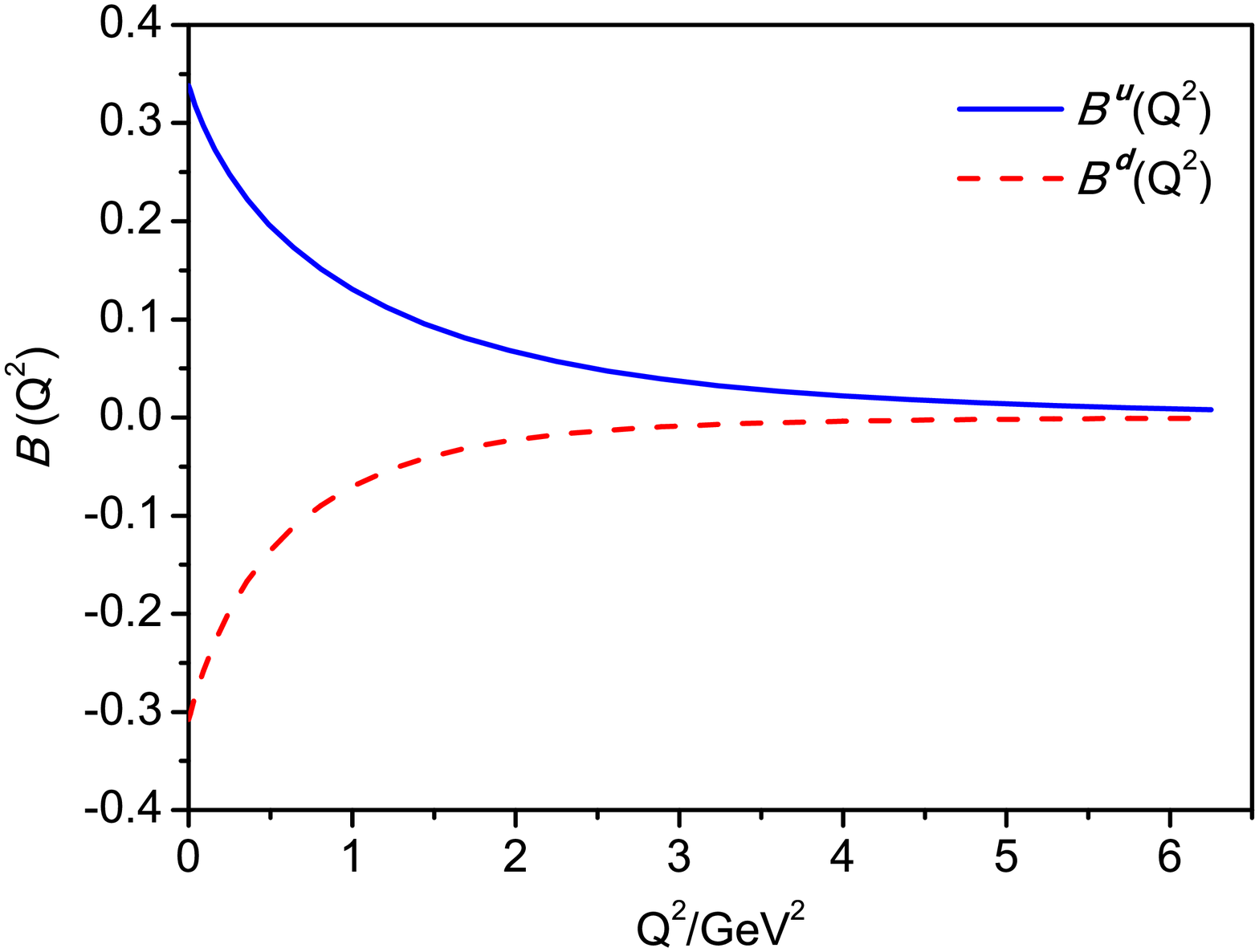}
\caption{(Color online). The generalized form factors $A(Q^2)$ plotted in the upper panel and $B(Q^2)$ plotted in the lower panel in the light-cone spectator-diquark model. The solid curve stands for the $u$ quark part, and the dashed curve stands for the $d$ quark part.\label{ngffv}}
\end{figure}

\begin{figure}
\includegraphics[width=0.5\textwidth]{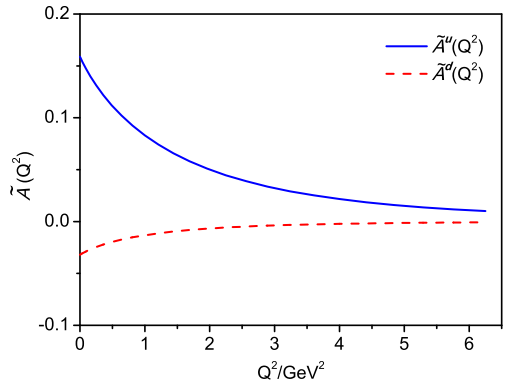}
\caption{(Color online). The generalized form factor $\tilde{A}(Q^2)$ in the light-cone spectator-diquark model. The solid curve stands for the $u$ quark part, and the dashed curve stands for the $d$ quark part.\label{ngffa}}
\end{figure}

\begin{figure}
\includegraphics[width=0.5\textwidth]{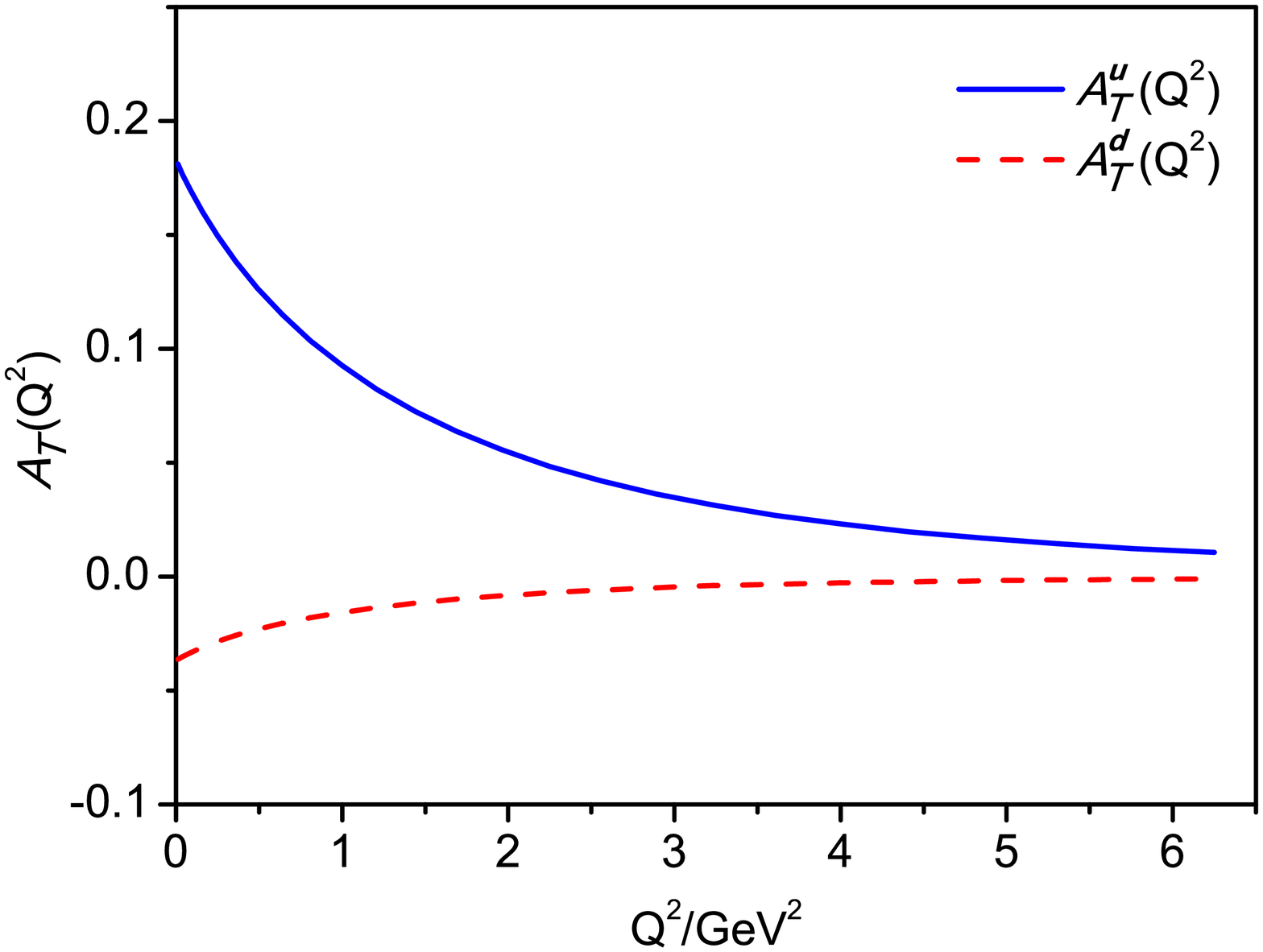}
\includegraphics[width=0.5\textwidth]{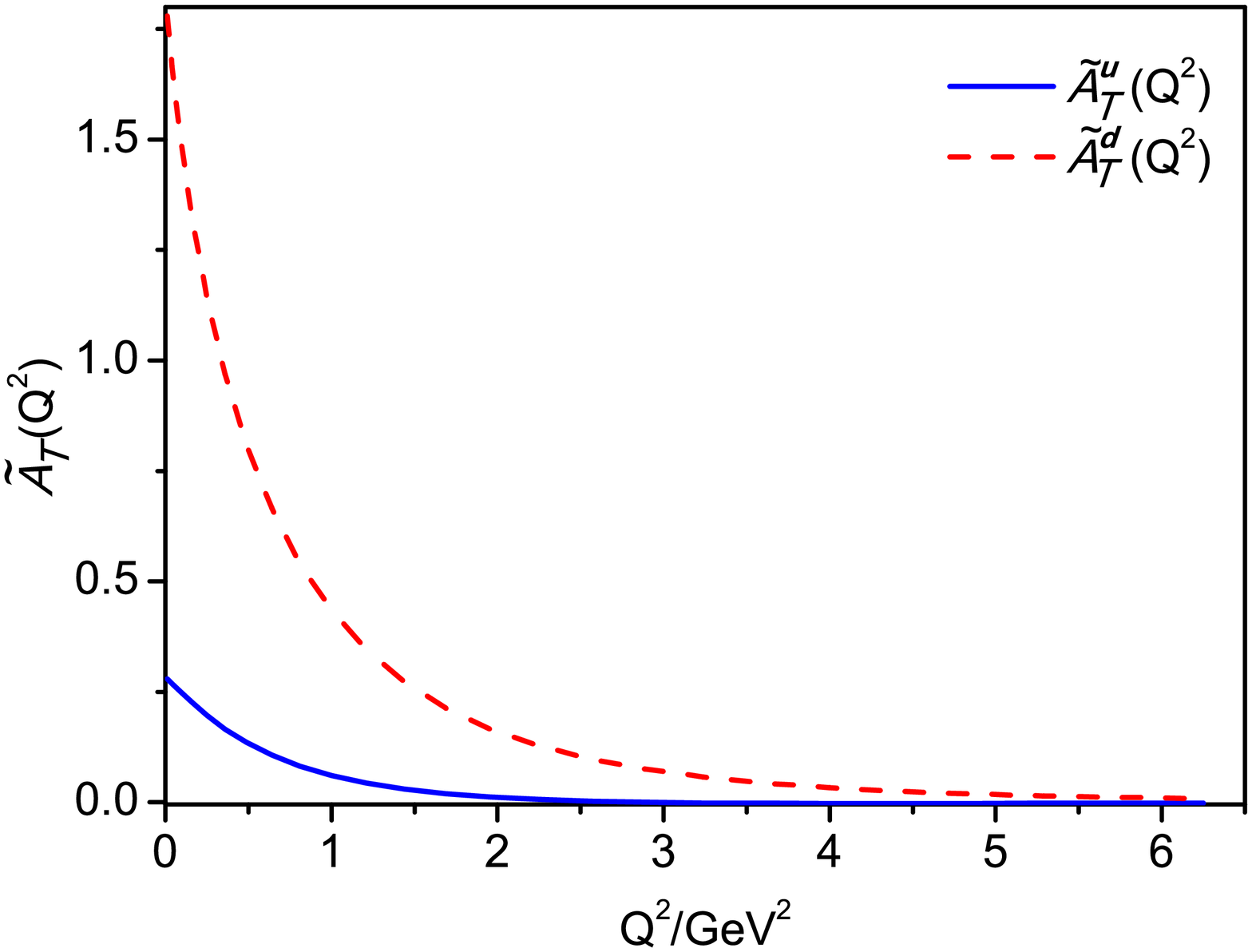}
\includegraphics[width=0.5\textwidth]{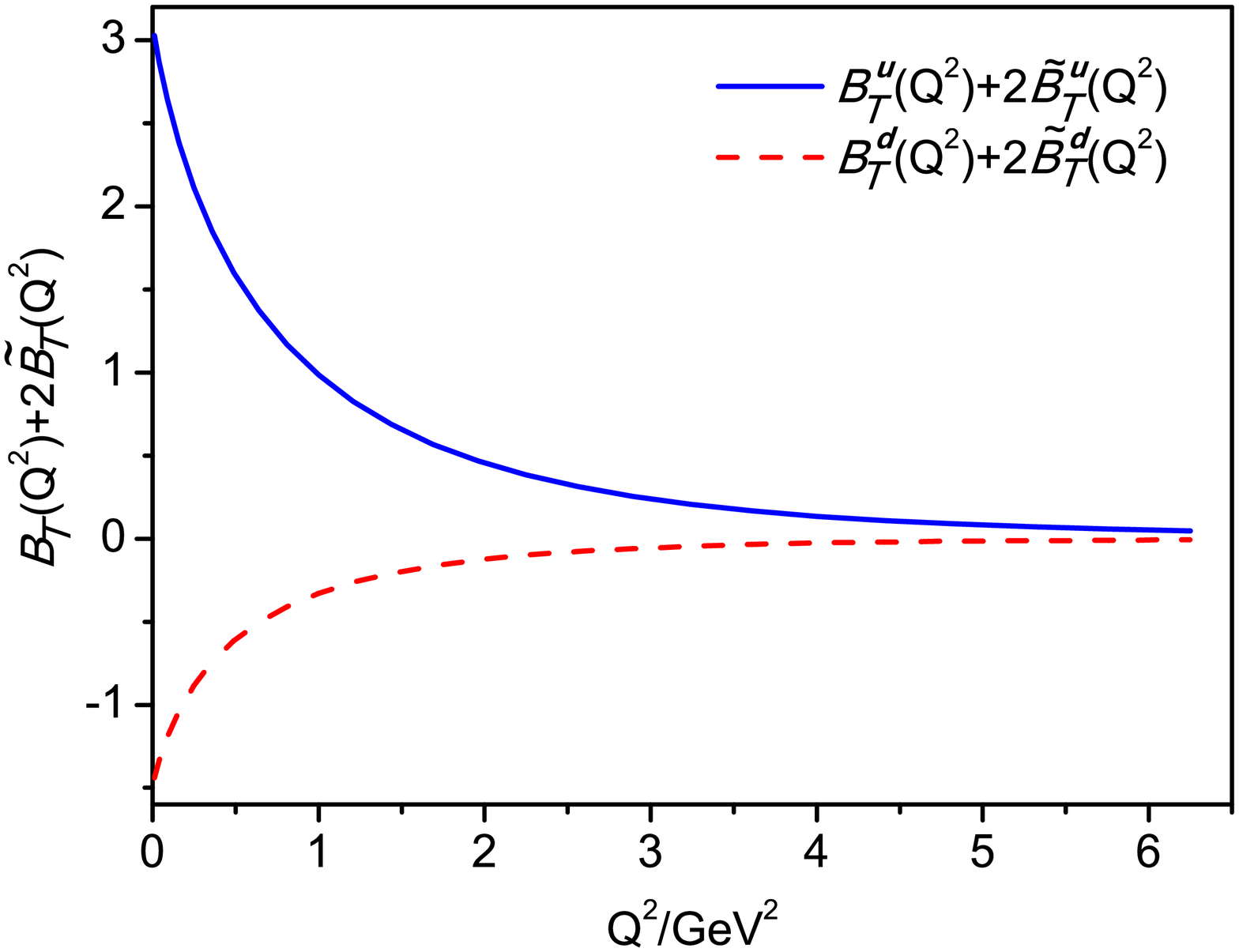}
\caption{(Color online). The generalized form factors $A_T(Q^2)$ plotted in the upper panel, $\tilde{A}_T(Q^2)$ plotted in the middle panel and $B_T(Q^2)+2\tilde{B}_T(Q^2)$ plotted in the lower panel in the light-cone spectator-diquark model. The solid curve stands for the $u$ quark part, and the dashed curve stands for the $d$ quark part.\label{ngfft}}
\end{figure}

We choose the parameters in the model for our numerical calculations as
\begin{equation}
\begin{split}
m=&330\,\textrm{MeV},\quad m_S=600\,\textrm{MeV},\quad m_V=800\,\textrm{MeV},\\
&\beta_S=\beta_V=330\,\textrm{MeV},
\end{split}
\end{equation}
which are similar as the choice in~\cite{Ma:1996np}. The difference of the masses of the scalar and axial-vector spectator-diquark is due to the spin interaction from color magnetism or alternatively from instantons~\cite{Weber:1992ww,Weber:1994sq}, and the values are chosen as estimated to explain the $N-\Delta$ mass difference. Phenomenologically, the calculated results with these values are in reasonable agreement with the experimental data, such as the ratio of structure functions $F^n_2(x)/F^p_2(x)$~\cite{Arneodo:1994sh}, the quark distributions~\cite{Ma:1996np} and the form factors~\cite{Ma:2002ir}. As mentioned in Sec. II, we choose the BHL prescription~\cite{Brodsky:1980vj,Brodsky:1981jv,Brodsky:1982nx} for the space part of the nucleon light-cone wave function. Instead of this, there are some alternative choices, such as the Teren'ev-Karmanov (TK) prescription~\cite{Terentev:1976jk,Karmanov:1979if} and the Chung-Coester-Polyzou (CCP) prescription~\cite{Chung:1988mu}. Besides, the light-cone wave function may also be solved through the AdS/CFT correspondence between the string states in anti-de Sitter (AdS) space and conformal field theories (CFT) in physical space-time as a first approximation to the QCD~\cite{Brodsky:2003px,Brodsky:2006uqa,deTeramond:2008ht}. To better describe the quark distributions on the light-cone momentum fraction $x$, we normalize the unpolarized $\bm{k}_\perp$-integrated parton distributions to the MSTW parametrizations~\cite{Martin:2009iq} by multiplying the light-cone wave function with a factor:
\begin{equation}
f_{_\textrm{MSTW}}^q(x)^{\frac{1}{2}}\Big(\int d^2\bm{k}_\perp f_1^q(x,\bm{k}_\perp)\Big)^{-\frac{1}{2}},
\end{equation}
where the $f_{_\textrm{MSTW}}^q(x)$ is the MSTW2008LO parametrizations and the $f_1^q(x,\bm{k}_\perp)$ is the unpolarized TMD calculated with the BHL wave function. The numerical results of the generalized form factors are plotted in Figs.~\ref{ngffv}-\ref{ngfft}.

In addition, the operator $\mathcal{O}^{\mu\nu}$ is just the quark part of the energy-momentum tensor:
\begin{equation}
\begin{split}
\Theta^{\mu\nu}=&i\bar{\psi}\gamma^{(\mu}\mathcal{D}^{\nu)}\psi-g^{\mu\nu}\bar{\psi}(i\slashed{\mathcal{D}}-m)\psi\\
&-F^{a\mu\rho}F^{a\nu}_{\ \ \rho}+\frac{1}{4}g^{\mu\nu}F^{a\rho\sigma}F^a_{\rho\sigma},
\end{split}
\end{equation}
which is obtained by varying the QCD action with respect to the space-time metric $g_{\mu\nu}$, and the plus-plus component of the second term vanishes. Thus the corresponding form factors are usually named as the energy-momentum tensor or the gravitational form factors. The energy-momentum tensor current is a spin-2 current and in principle couples to the graviton. Thus the gravitational form factors are ``measurable'' via elastic graviton proton scatterings as mentioned in some literature~\cite{Brodsky:2000ii}, but this experiment is infeasible at least at present. Since the energy-momentum tensor operator does appear in the operator product expansion for a product of two vector currents $T\{j^\mu(\xi)j^\nu(0)\}$, it is suggested to measure the $A(0)$ and $B(0)$  through the deeply virtual Compton scattering (DVCS) process~\cite{Ji:1996ek} and the deeply virtual meson production (DVMP) process. However, to extract the $Q^2$ dependence of these form factors is quite a challenging issue, and there are no efficient methods yet. Besides, the generalized form factors can also be obtained via the Mellin moments of the generalized parton distributions (GPDs) which provide a comprehensive framework for describing the parton structure of the nucleon:
\begin{equation}
\int_{-1}^1dx\,xG(x,\xi,Q^2)=F(Q^2),
\end{equation}
where $G$ represents the GPDs $H$, $E$, $\tilde{H}$, $\tilde{E}$, $H_T$, $E_T$, $\tilde{H}_T$ and $\tilde{E}_T$, and $F(Q^2)$ represents the $n=2$ generalized form factors $A(Q^2)+4\xi^2C(Q^2)$, $B(Q^2)-4\xi^2C(Q^2)$, $\tilde{A}(Q^2)$, $\tilde{B}(Q^2)$, $A_T(Q^2)$, $B_T(Q^2)$, $\tilde{A}_T(Q^2)$ and $-2\xi\tilde{B}_T(Q^2)$ respectively. There are similar relations between the $n$-th Mellin moments of the GPDs and the generalized form factors for $n>2$ cases.

\begin{figure}
\includegraphics[width=0.5\textwidth]{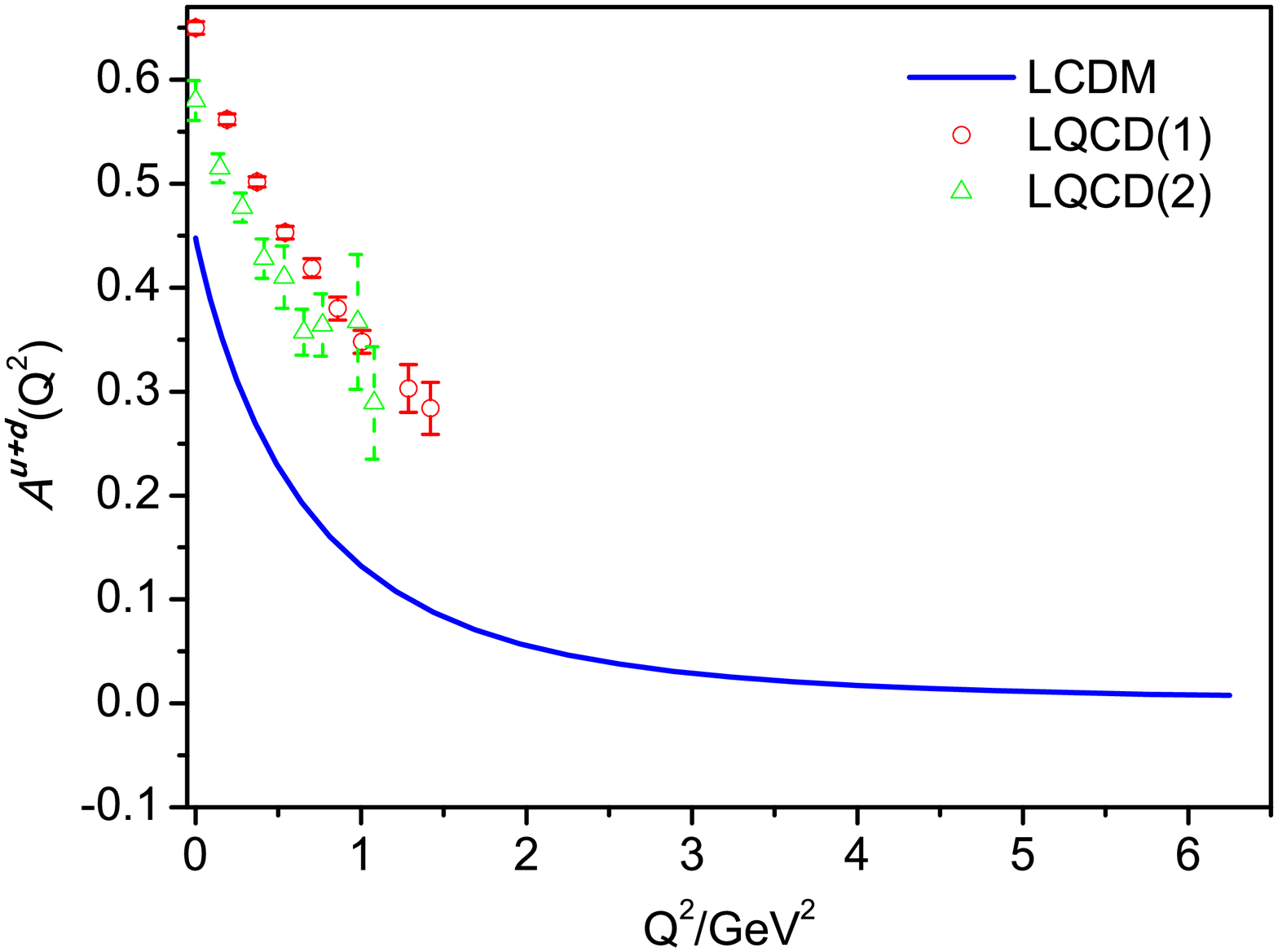}
\includegraphics[width=0.5\textwidth]{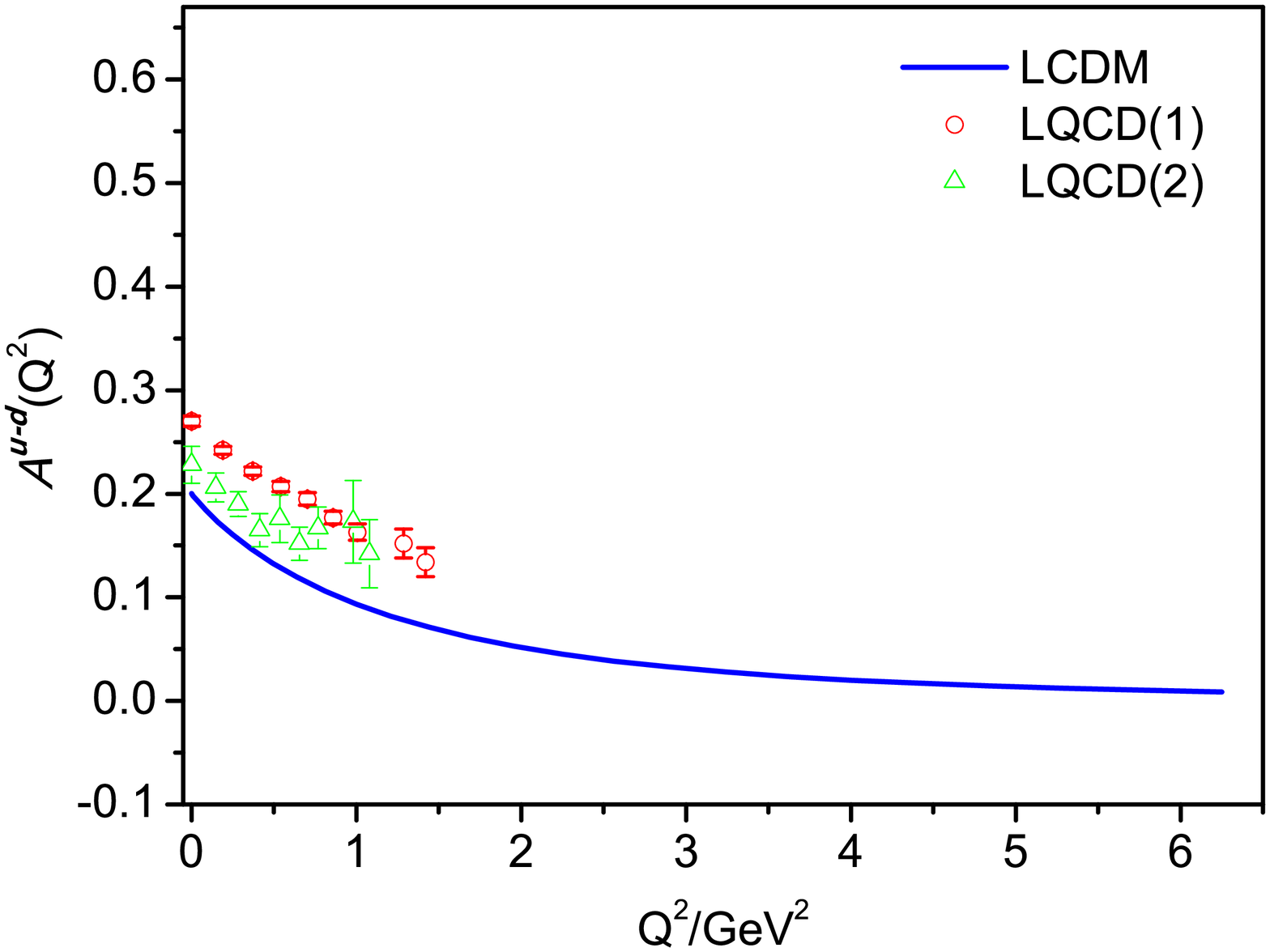}
\caption{(Color online). The $A(Q^2)$ in the light-cone spectator model (LCDM) compared with some lattice QCD (LQCD) results. The isoscalar $(u+d)$ part is plotted in the upper panel and the isovector $(u-d)$ part is plotted in the lower panel. The solid curve stands for our LCDM results. The circle markers stand for the LQCD results in Ref. \cite{Alexandrou:2013joa} with pion mass 375\,MeV and the triangle markers stand for those with pion mass 213\,MeV.\label{ascalar}}
\end{figure}

\begin{figure}
\includegraphics[width=0.5\textwidth]{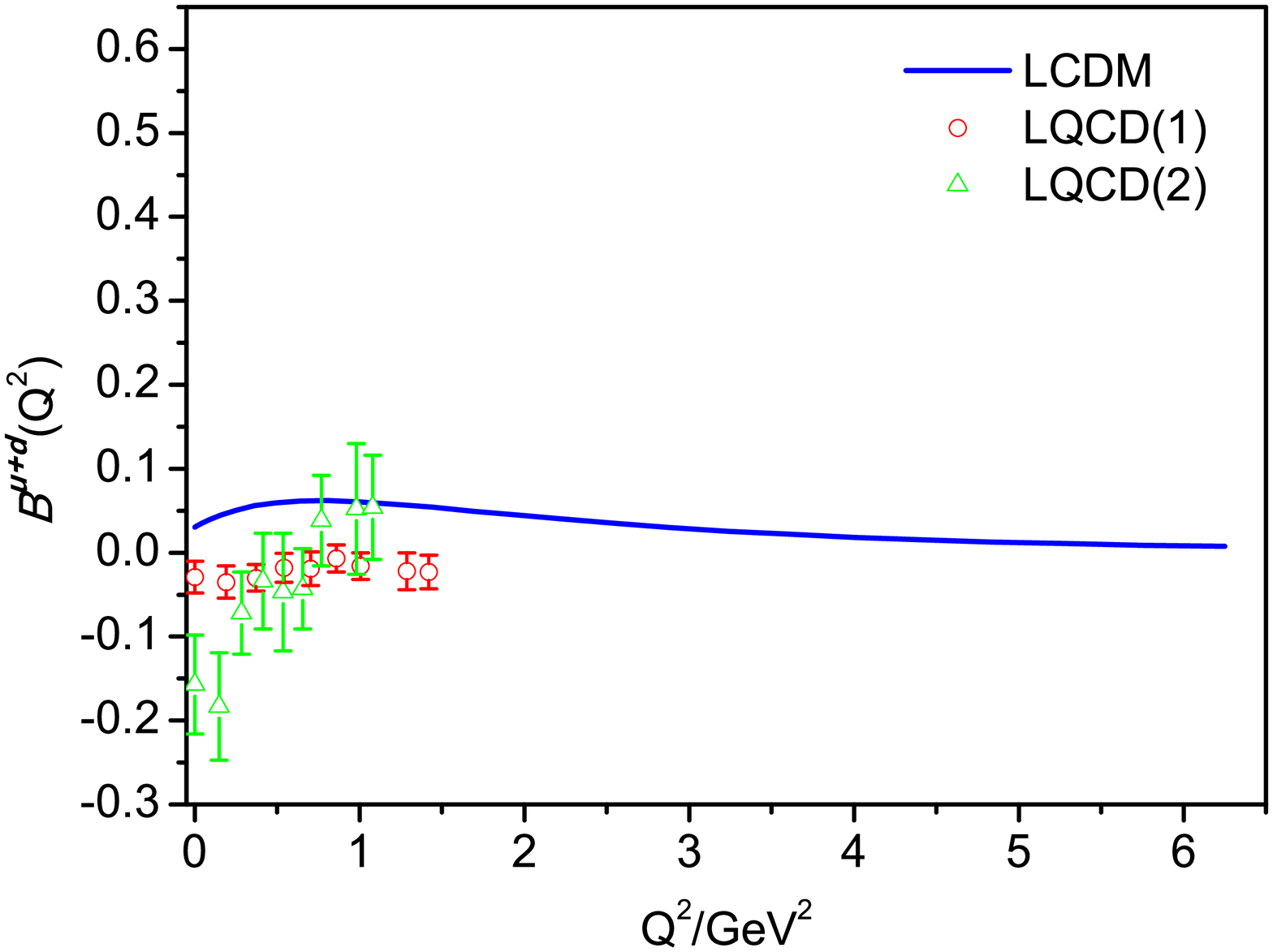}
\includegraphics[width=0.5\textwidth]{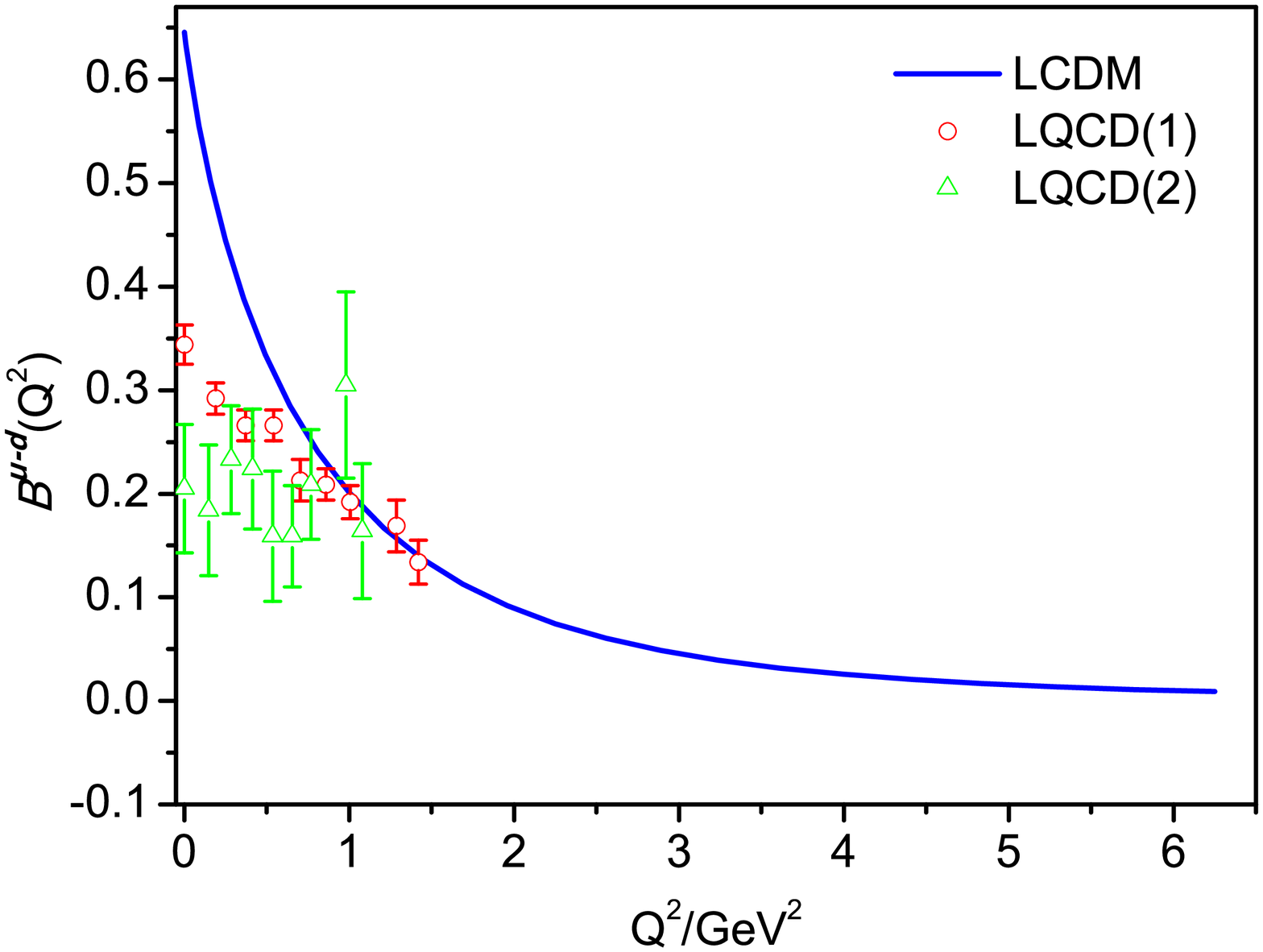}
\caption{(Color online). The $B(Q^2)$ in the light-cone spectator model (LCDM) compared with some lattice QCD (LQCD) results. The isoscalar $(u+d)$ part is plotted in the upper panel and the isovector $(u-d)$ part is plotted in the lower panel. The solid curve stands for our LCDM results. The circle markers stand for the LQCD results in Ref. \cite{Alexandrou:2013joa} with pion mass 375\,MeV and the triangle markers stand for those with pion mass 213\,MeV.\label{bscalar}}
\end{figure}

\begin{figure}
\includegraphics[width=0.5\textwidth]{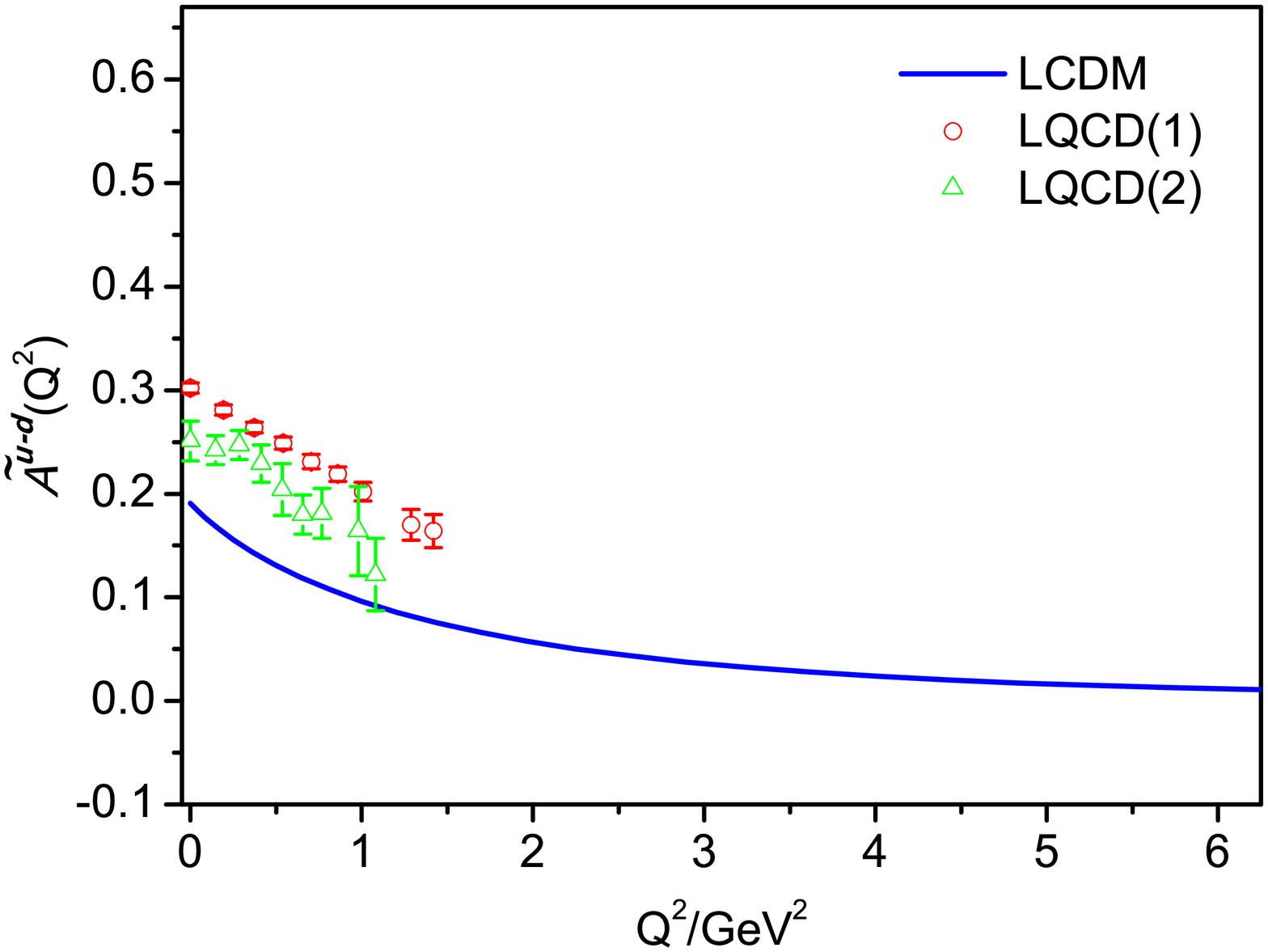}
\caption{(Color online). The isovector $\tilde{A}(Q^2)$ in the light-cone spectator model (LCDM) compared with some lattice QCD (LQCD) results. The solid curve stands for our LCDM results. The circle markers stand for the LQCD results in Ref. \cite{Alexandrou:2013joa} with pion mass 375\,MeV and the triangle markers stand for those with pion mass 213\,MeV.\label{a5vector}}
\end{figure}

The DVCS process was measured at the Jefferson Lab~\cite{MunozCamacho:2006hx,Mazouz:2007aa,Girod:2007aa} and at HERA~\cite{Chekanov:2003ya,Aktas:2005ty,Airapetian:2006zr,Aaron:2007ab,Chekanov:2008vy}. Some information of the GPDs is already obtained, but it is still a big challenge to extract the GPDs precisely. The future experiments with high luminosity and resolution, such as the COMPASS-II~\cite{Chiosso:2013ila,Silva:2013dta} and the Jefferson Lab 12 GeV experiment~\cite{Roche:2012pr}, will improve the measurement. However, the analysis for extracting the GPDs requires also a partial modeling of the combined dependence on the momentum fraction $x$, the skewness parameter $\xi$ and the virtuality $t=-Q^2$. An alternative method for cross-checks is given by the lattice QCD. Although the GPDs are not directly accessible on the lattice at present, their Mellin moments, {\it i.e.} the generalized form factors, are accessible. We compare our calculations with some lattice QCD results~\cite{Alexandrou:2013joa} in Figs. \ref{ascalar}-\ref{a5vector}. Whereas, the results from the lattice QCD still have much dependence on the parameters. For example, the pion mass, as an important parameter in lattice computations, cannot be chosen as small as the real physical value because of the limitation on the computational techniques~\cite{Sternbeck:2012rw,Bali:2013dpa}. In addition, as shown in~\cite{Bali:2013dpa}, the optimizations in lattice computations may also cause visible changes. Therefore, the results from lattice QCD still have large uncertainties.

\section{Summary}

In this paper, we have calculated the $n=2$ generalized form factors of the nucleon in a light-cone spectator-diquark model. The Melosh-Wigner rotation effect is included for both the quark and the axial-vector diquark. This model is proved successful in some phenomenology investigations, such as form factors, structure functions and TMDs. As a further test of the model and an investigation of the structure of the nucleon, it is significant to extend its application to other observables. Compared to the ordinary form factors, {\it i.e.} the $n=1$ case, the generalized form factors contain much richer information of the nucleon. Experimentally, the generalized form factors can be measured via the DVCS or DVMP processes, since they are Mellin moments of the GPDs which provide a three-dimensional picture of the nucleon. On the other hand, the generalized form factors can reflect the properties of the GPDs. However, extracting the GPDs from experiments is quite challenging and partially model dependent. As an alternative method for cross-checks, the generalized form factors are accessible on the lattice QCD. We also compared our results with some lattice QCD data. By taking the uncertainties in lattice computations into account, our results are comparable with the lattice QCD data. Therefore, this study will help us to understand the three-dimensional parton structure of the nucleon as well as the nonperturbative QCD properties.


\begin{acknowledgments}
This work is supported by the National Natural Science Foundation of China (Grants No. 11021092, No. 10975003, No. 11035003 and No. 11120101004).
\end{acknowledgments}


%

\end{document}